\documentstyle[12pt,aasms4]{article}

\lefthead{Antiochos, McNeice, Spicer, \& Klimchuk}
\righthead{Dynamic Formation of Condensations}

\begin{document}

\title{The Dynamic Formation of Prominence Condensations}

\author{S. K. Antiochos}
\affil{E.O. Hulburt Center for Space Research, Naval Research Laboratory,
    Washington, DC, 20375}

\author{P. J. MacNeice}
\affil{Raytheon STX Corporation, Greenbelt, MD, 20770}

\author{D. S. Spicer}
\affil{Earth and Space Data and Computing Division, NASA/Goddard Space 
Flight Center, Greenbelt, MD, 20771}

\and

\author{J. A. Klimchuk}
\affil{E.O. Hulburt Center for Space Research, Naval Research Laboratory,
    Washington, DC, 20375}

\begin{abstract}

We present simulations of a model for the formation of a prominence
condensation in a coronal loop. The key idea behind the model is that
the spatial localization of loop heating near the chromosphere leads
to a catastrophic cooling in the corona (\cite{Antiochos91}).  Using a
new adaptive grid code, we simulate the complete growth of a
condensation, and find that after $\sim 5,000$ s it reaches a
quasi-steady state. We show that the size and the growth time of the
condensation are in good agreement with data, and discuss the
implications of the model for coronal heating and SOHO/TRACE
observations.

\end{abstract}


\section{Introduction}
Solar prominences and filaments are beautiful phenomena both in their
appearance and in their physics.  They are observed in H$\alpha$ as
large masses $\sim 10^{16}$g of dense $> 10^{11}$ cm$^{-3}$
chromospheric plasma which seem to float high up in the tenuous
($\sim10^{9}$ cm$^{-3}$) solar atmosphere, the million-degree
corona. They can have a myriad of shapes, but often resemble a long
arched wall with widths ranging from $10^8$ to $10^9$ cm, heights
ranging from $10^{10}$ cm down to the chromosphere itself, and lengths
that can be of order the Sun's circumference for high-latitude polar
crown filaments. Typical lifetimes of prominences can range from as
short as hours for active region filaments to as long as months for
quiescent prominences.  Prominences usually disappear by
eruption. Large coronal mass ejections and two-ribbon flares are
almost always accompanied by prominence eruption.  Consequently, the
structure of prominences, especially their magnetic structure, is
vital for understanding the major solar drivers of geomagnetic
disturbances.

There are three central issues concerning the physics of prominences:
How does mass condense in the corona, how is it supported, and why
does it erupt? The coronal magnetic field undoubtably plays a vital
role in all three issues ({\it e.g.} \cite{Hanssen95,Priest89}).  We
believe, however, that these questions are essentially decoupled in
that the eruption mechanism is due to the global topology of the field
(\cite{Antiochos98,Antiochos99}), the support is due to the local
geometry of a sheared 3D arcade (\cite{Antiochos94}), while the
condensation formation is due to the properties of the coronal heating
process and the cooling of coronal plasma by radiation and conduction
(\cite{Antiochos91,Dahlburg98}). In this paper we focus on the third
question of condensation formation and, in particular, address the
dynamics of this process.

When viewed with high spatial resolution, it can be seen that the mass
of a prominence is not one continuous structure, but is actually a
collection of small condensations or knots that form on time scales of
tens of minutes and have size scales of order a thousand
km. (\cite{Engvold76,Zirker89}). Given typical coronal gas pressures,
$\sim 10^{-1}$ ergs/cm$^3$ and magnetic field strengths 10 G, the
plasma is low-beta and we expect that each condensation is isolated
from the others by the magnetic field. Therefore, the canonical
picture of a prominence is that of an arcade of flux tubes, or coronal
loops, each of which has somewhere along its length a plug of cool
material that is observed as an H$\alpha$ knot, and the ensemble of
all the knots forms the global structure that we call a prominence.

An important point is that prominence knots cannot be considered as
simply the condensation of coronal plasma because they generally
contain more mass than the coronal portion of the flux tubes they
occupy. Material must be brought up from the chromosphere in order to
produce the large mass of a prominence, which has led many authors to
consider the so-called siphon flow model for condensation formation
({\it e.g.}
\cite{Pikelner71,Engvold77,Priest79,Ribes80,Poland86,An88,Mok90}).
It has proved to be difficult, however, for such models to produce
condensations of sufficient mass (\cite{Poland86}).  The observation
that the knot contains a large mass implies a large amount of
chromospheric evaporation ({\it e.g.}  \cite{Antiochos78}) and, hence,
a large coronal heating rate in the loop. But a high heating rate
implies high coronal-like temperatures, rather than the cool
chromospheric temperatures needed for a prominence.

We have described elsewhere how one can reconcile the seemingly
contradictory requirements of high coronal heating and low coronal
temperatures (\cite{Antiochos91,Dahlburg98}). The key idea is that if
the heating in a coronal loop has a strong spatial dependence, then it
is possible to have a high heating rate and still develop a massive
condensation near the loop center. Note that if the heating is uniform
per unit volume or per unit mass, which are the usual assumptions in
coronal loop models ({\it e.g.}  \cite{Rosner78,Vesecky79}), then the
loop can only have a hot structure with coronal temperatures
throughout. This is true even if the loop has a dipped geometry
suitable for supporting prominence material (\cite{Antiochos91}). If
the heating is sufficiently localized near the loop base, however, the
plasma structure can change to one in which there are chromospheric
temperatures in the dipped portion of the loop.

In our previous paper ({\cite{Antiochos91}), we verified this
hypothesis with numerical simulation.  We began the simulation with a
coronal loop that had a dipped geometry and a spatially uniform
heating. This loop developed a typical hot coronal equilibrium
solution with a temperature maximum $> 10^6$K at the loop midpoint and
a thin transition region at the base. We then {\it increased} the
heating, but only in a small region near the base.  At first, this
heating increase caused the temperature to rise throughout the loop,
so that the central dipped portion became even hotter.  Chromospheric
evaporation was driven by the temperature rise, which caused the
density and, consequently, the radiative losses to increase throughout
the loop, even in the upper section where the heating was
unchanged. Eventually, the increase in radiative losses produced a
catastrophic cooling at the loop center, and the formation of a
chromospheric condensation.

It is worth noting that this temperature collapse is physically due to
a loss-of-equilibrium rather than a thermal instability. For small
increases of the heating near the base, it is possible to find static
equilibrium solutions which have a hot corona throughout the loop.
These solutions are characterised by a slight increase in the
temperature at the loop midpoint as the heating increases.  But if the
base heating is increased sufficiently, then no static solution with a
hot midpoint is possible, because that would imply too high a
radiative loss rate there.  The only static solutions are those in
which the temperature in the middle section of the loop is very low,
$ T << 10^5$ K, so that the radiative losses there are reduced
(\cite{Dahlburg98}).

Although our previous simulation verified the basic idea of the model,
there was a problem with that calculation which prevented us from
comparing the model with observations, especially the time scale for
prominence formation. The code could not follow the condensation
significantly into its growth phase.  As the condensation formed, a
transition region with very large temperature gradients formed on each
side of the condensation, exactly like the transition region at the
footpoints. This transition region moved down the loop as the
condensation grew, but since our code used a fixed grid, it was not
possible to resolve a moving transition region and, consequently, the
simulation broke down almost immediately after the condensation first
appeared (\cite{Antiochos91}). In this paper we present simulations
using a new code that has a fully adaptive grid (\cite{MacNeice98})
and, therefore, can follow a condensation throughout its evolution,
allowing us to compare the growth times with data. These new
calculations also include more realistic forms for the loop geometry
and the heating.

\section{The Numerical Model}

Since the magnetic field in a prominence dominates the plasma, we make
the usual assumption of a 1D coronal loop model in which the field
determines the loop geometry and all plasma quantities vary only
along the field. Our code solves the standard set of transport
equations for a 1D plasma:

\begin{equation}
\frac{\partial}{\partial t} \rho + \frac{\partial}{\partial s} (\rho V) = 0,
\end{equation}
\begin{equation}
\frac{\partial}{\partial t}(\rho V) + \frac{\partial}{\partial s}
(\rho V^2 + P) = - \rho  g_{\|}(s),
\end{equation}
and
\begin{equation}
\frac{\partial}{\partial t}U + \frac{\partial}{\partial s}(U V 
- 10^{-6} T^{5/2}\frac{\partial}{\partial s} T) 
= - P\frac{\partial}{\partial s}V + E(t,s) - n^2 \Lambda (T).
\end{equation}
In Equations (1) -- (3), $s$ is distance along the loop, {\it i.e.}
along the magnetic flux tube; $\rho$ is the plasma mass density; $V$
is the velocity along the loop; $P$ is gas pressure; $g_{\|}$ is the
component of gravity along the field; $U = 3 P /2 $ is 
the internal energy;
$10^{-6} T^{5/2}$ is the Spitzer
(1962) coefficient of thermal conduction; $E(t,s)$ is the assumed form
for the coronal heating rate; $n$ is the electron number density; and
$ \Lambda (T)$ is the radiative loss coefficient for optically thin
emission ({\it e.g.} \cite{Cook89}). Note that we assume a fully
ionized hydrogen plasma, so that $\rho = 1.67 \times 10^{-24}\, n$
and the equation of state is simply $P = 2n k T$.

The code that we have developed (\cite{MacNeice98a}) to solve these
equations has two major improvements over our previous code. First, in
order to advance the solution, it uses a second order Godunov scheme
with a MUSCL limiter (\cite{VanLeer79}) applied to the characteristic
variables.  It is widely believed that higher order Godunov schemes,
which solve the nonlinear Riemann problem at each cell face, are the
most robust methods for numerically solving 1D fluid flow problems
like the one we expect for coronal loop plasma --- a non-turbulent
compressible flow in which steep gradients and/or discontinuities are
likely to develop (\cite{Hirsch88,Colella90}).

Second, the code employs a fully adaptive grid by using the PARAMESH
Adaptive Mesh Refinement package (\cite{MacNeice98}). PARAMESH builds
a hierarchy of sub-grids to cover the computational domain, with
spatial resolution varying to satisfy the demands of the application.
These sub-grid blocks form the nodes of a tree data-structure.  Each
grid block has a logically cartesian mesh, and the index ranges are
the same for every block. In this case we use 1D sub-grid blocks with
20 grid cells and 2 guard cells at each end of the sub-grid. When a
sub-grid block needs to be refined it spawns 2 child blocks. These two
children have their own 20 cell sub-grids, and together cover the same
coordinate range as their parent, but now with mesh spacing one-half
that of the parent's mesh spacing. Any or all of these children can
themselves be refined in the same manner. PARAMESH provides routines
which manage the necessary communication between sub-grid blocks.

An adaptive grid is essential for our simulation because we expect a
thin transition region to form somewhere in the loop and to move with
time. Unless this region is resolved, the plasma energetics and
dynamics cannot be calculated accurately. Often, a tricky issue is the
selection of a criterion for refining and derefining the numerical
grid. We find that density is the best quantity to use in determing
the refinement because it exhibits a strong spatial variation, both in
the transition region where pressure is approximately constant, and in
the chromosphere, where temperature is approximately constant. In the
simulations presented here we use six levels of refinement, which
corresponds to a minimum grid spacing of $\sim 5$ km. This is
sufficient to do an adequate job of resolving the transition region
throughout the evolution.

For these calculations we established a very simple
refinement/derefinement criterion. At each grid cell $i$ 
we computed the error measure
\begin{equation}
\epsilon_i = \vert \rho_i - \rho_{i-1} \vert /
{\rm min}( \rho_i , \rho_{i-1}),
\end{equation}
and for each sub-grid found the maximum value
\begin{equation}
\epsilon_{max} = {\max} ( \epsilon_0, \epsilon_1, \ldots, \epsilon_{21}).
\end{equation}
Notice we include one guard cell at each end of the sub-grid (ie $i=0, 
i=21$) to enable refinement in advance of an arriving feature of the
solution.  If $\epsilon_{max}$ exceeds 0.25, meaning that somewhere in
the sub-grid block the density varies by at least 25\% between grid
cells, then that block is refined. If $\epsilon_{max}$ is less than
0.05 on a pair of sibling sub-grid blocks, then the block is derefined
provided the lower resolution representation of their parents does not
exceed the refinement threshold.

The refinement process was permitted to range freely between
refinement levels 6 and 12. In this context, level 1 indicates that
the entire computational domain is covered by 1 sub-grid block. A
level 1 sub-grid would have a cell size of $\Delta x = L/20$ where $L$
is the total length of the computational domain.  A level $n$ sub-grid
therefore has a cell size of $2^{-n+1}L/20$.  The initial grid was
uniformly refined at level 6, ie $\Delta x = 2^{-5}L/20 = 0.0015625
L$.  The sub-grids refined to level 12 have a cell size which is a
factor $2^6 = 64$ finer.

The geometry that we assume for the loop is shown in Figure 1. For
simplicity, we took the loop to be symmetric about its midpoint.  In a
1D model the loop (i.e. magnetic field) geometry influences the plasma
evolution solely through its effect on gravity. The only significant
restriction on the magnetic geometry is that the flux tube have a
dipped section, as shown in Figure 1, which can support material
against gravity. The geometry of Figure 1 would be appropriate for any
of the usual prominence magnetic field models such as the
Kippenhahn-Schluter (1957) or Kuperus-Raduu (1974), or even our recent
3D model (\cite{Antiochos94}).

The loop is comprised of three distinct sections. First, is
a straight vertical tube of height $s_1$ --- this is roughly the
chromospheric section. The exact position of the top of the
chromosphere changes during the calculation.  This section of the
loop ($s \le s_1$) is given by
\begin{equation}
z(s) = s, 
\quad \hbox{hence} \quad 
g_{\|}(s) = g_{\odot},
\end{equation}
where $z(s)$ is height at position $s$, and $g_{\odot} = 2.7 \times
10^4$ cm/s$^2$ is the solar gravity.  Next is a quarter circle of arc
length $s_2 - s_1$ and, therefore, radius $R = 2 (s_2 - s_1)/
\pi$. Note that the maximum height of the loop is $s_1 + R$. This
section ($s_1 < s \le s_2 $) is given by,
\begin{equation}
z(s) = s_1 + R \sin(\frac{s-s_1}{R}), 
\quad \hbox{hence} \quad
g_{\|}(s)= g_{\odot} \cos(\frac{\pi}{2} \frac{s - s_1}{s_2 - s_1}) .
\end{equation}
The final section is a quasi-cosine curve that has a maximum height at $s_2$ to
match the quarter circle, and a minimum height at the loop midpoint,
$s = L$, and has a total depth $D$. This part ($s_2 < s \le L$) is given by:
\begin{equation}
z(s) = s_1 + R - \frac{D}{2} \left(1 - \cos(\pi \frac{s - s_2}{L - s_2})\right),
\quad \hbox{hence} \quad
g_{\|}(s)= -g_{\odot} \frac{\pi D}{2(L - s_2)} \sin(\pi \frac{s - s_2}{L - s_2}).
\end{equation}
>From the form above, we note that the gravity term in the momentum
equation is negative for $s \le s_2$, and positive for $s > s_2$. The
gravity term is continuous, but does not have continuous
derivatives.

The geometry parameters that we must specify (Figure 1) are the
position of the top of the chromosphere, $s_1 = 1 \times 10^9$ cm, the
position of the loop apex $s_2 = s_1 + 2.5 \times 10^9$ cm (therefore
the maximum loop height $s_1 + R = s_1 + 5 \times 10^9 / \pi$ cm); the
total loop half-length, $L = 11 \times 10^9$ cm, and the depth of the
coronal dip, $D = 5 \times 10^8$cm. These values are much more
appropriate for quiescent prominence loops than the values used in
our previous work, which are suitable only for low-lying active region
prominences (\cite{Dahlburg98}).

For the radiative losses we take a simple analytic form.  Since the
radiative losses depend sensitively on elemental abundances which vary
substantially on the Sun (\cite{Cook89}), there is no longer a
``standard'' curve to use. The important point is that our model does
not depend on the details of the radiative loss curve; it is sensitive
only to the spatial dependence of the heating.  In the upper
transition region and corona ($T \ge 10^5$ K), we take $ \Lambda(T) =
1.0 \times 10^{-17} / T$.  In the lower transition region ($ 10^5 > T
\ge 30,000$ K), $ \Lambda(T) = 1.0 \times 10^{-37} T^3$.  A
temperature at the base of the model of 30,000 K is assumed. We would
like the chromospheric region of the loop to remain at this
temperature as much as possible; consequently, we take the radiative
losses to vanish below 29,500 K. For $30,000 > T > 29,500$ K,
$\Lambda(T) = 2.7 \times 10^{-24} ((T - 29,500)/500)$, and $\Lambda(T)
= 0.0$ for $ T \le 29,500$ K. In addition, we drop the density
dependence of the radiative losses in the chromospheric region where
the density can become extremely large due to the exponential increase
per scale height.  In the Sun's chromosphere the radiative losses are
limited by radiative transfer, which is not included in our model. As
with $g_{\|}(s)$ the assumed form for $\Lambda(T)$ is continuous, but
without continuous derivatives.

Finally, the form of the heating needs to be specified.  We set $E =
E_0 + E_1(s,t)$, where $E_0 = 1.5 \times 10^{-5}$ ergs/cm$^3$/s, is
the uniform background heating which stays on throughout the
simulation. The spatially dependent heating $E_1$ is ramped up over
1000 s only after the loop has settled into a static equilibrium with
the uniform heating. We set $E_1$ to be constant in the chromosphere,
$E_1 = 10^{-3}$ ergs/cm$^3$/s, for $s \le s_1$, and have an
exponential drop-off in the corona,
\begin{equation}
E_1 = 10^{-3} \exp(-(s - s_1)/ \lambda), \quad \hbox{for} \quad s > s_1,
\end{equation}
where the damping length $\lambda$ is chosen to be 10,000 km.
This form for the spatial dependence of the heating is much
more physical than that used previously (\cite{Antiochos91}), in which
we assumed that the heating was concentrated in a very narrow region
centered at the highest point in the loop. Here, there is no special
relation between the heating and the loop geometry. We are simply
assuming that the heating propagates into the loop from below, and
is deposited over some spatial scale small compared to the loop length.

\section{Results}

In order to solve Equations (1) -- (3), appropriate boundary and
initial conditions must be specified.  As boundary conditions on the
loop we assumed a rigid, constant temperature base at $s = 0$, and
symmetry at the top, $s = L$.  To derive the initial equilibrium we
simply started with a discontinuous temperature profile in which $T =
10^6$ K in the corona $s \ge s_1$, and $T = 30,000$ K in the
chromosphere. The density was constant in the corona, $n = 0.7 \times
10^9$ cm$^{-3}$ for $s \ge s_1$, and increased exponentially with
depth in the chromosphere at the appropriate scale height. Hence, the
plasma was in approximate force balance to start with, but not in
thermal equilibrium. We then let the system evolve until an
equilibrium was reached.

After $\sim 35,000$ s the loop had settled into a static equilibrium
with negligible residual motions, $V_{max} < 0.1$ km/s. We took this
state to be the initial equilibrium, $t \equiv 0$ s. Figures 2, 3, and
4 show the temperature, density and velocity profiles respectively at
this time. Even though the loop has a dip geometry, the temperature
and density profiles are exactly what one would expect for a standard
coronal loop (\cite{Rosner78,Vesecky79}). The maximum temperature
occurs at the loop midpoint and has a value of $1.2 \times 10^6$ K.
Note also that the top of the chromosphere has moved only slightly
upward from its starting position at $s_1 = 10,000$ km.

We then turned on the spatially dependent heating, $E_1$, ramping it
up linearly over 1,000 s. The plasma responded as expected by first
heating and evaporating chromospheric material into the loop before a
condensation appeared at the midpoint.  This evaporative phase of the
evolution lasted a considerable time, approximately 60,000 s.  Figures
2, 3, and 4 show the plasma profiles at three times: 10, 30, and 50
thousand seconds after the spatial heating turn-on. The temperature at
the loop midpoint reached a maximum value greater than $1.7\times10^6$
K at around 10,000 s, but then started to decline as the radiation
losses at the midpoint began to exceed the heating there. The position
of the top of the chromosphere moved downward over 1,000 km as a
result of the temperature increase and accompanying pressure increase,
but then moved back to nearly its original position as the temperature
dropped.

We note from Figure 3 that the density throughout the loop increased
steadily during the evaporative phase, except near the very end ($t >
50,000$ s) when the condensation started to form at the midpoint. Due
to the temperature collapse there, the pressure decreased rapidly at
the midpoint, which tended to evacuate the loop leg. This effect can
be seen in the velocity profiles. At early times ($\sim 10,000$ s) the
velocity is driven by evaporation, i.e., an overpressure at the top of
the chromosphere, consequently, the velocity peaks ($\sim 5$ km/s)
near the loop base (Figure 3). The evaporative velocity decreased,
however, as the loop temperature decreased. Later in the evolution the
velocity started to increase again, but now peaking nearer the
midpoint, which indicates that it was driven by the underpressure in
the condensation.

Once the condensation appeared at the loop midpoint, the subsequent
evolution was dominated by its rapid growth.  The temperature at the
loop midpoint first dropped below the chromospheric value of 30,000 K
at $t=54,000$ s. Figures 5 and 6 show the temperature and density
profiles over the last 10,000 km of our half-loop at this time and at
four later times in the evolution, 500, 1,000, 10,000, and 30,000 s
after the condensation appearance.  The corresponding velocity
profiles over the whole loop are shown in Figure 7.

There were two distinct phases to the condensation evolution: a
transient-motion phase characterized by large upward and downward
flows, and then a quasi-steady state characterized by slow upflows
throughout the loop.  At first, the condensation grew very rapidly as
the middle section of the loop dropped down to the base temperature of
30,000 K. This temperature collapse and the resulting pressure
decrease was so fast that a weak shock wave was launched from the loop
midpoint towards the base, Figure 7.  It can be seen from the Figure
that the wave speed was 160 km/s, which agreed well with the sound
speed in the corona. Weak shocks, like ordinary acoustic waves, are
expected to travel at the sound speed (e.g. \cite{Zeldovich68}). The
maximum upward velocity of the coronal plasma was less than 60 km/s,
however, so the bulk motion was clearly subsonic and no strong shocks
developed. The wave hit the chromosphere and then bounced back up the
loop with a reduced amplitude, Figure 7.  As shown in Fig. 7, the loop
oscillated back and forth for several bounces with an oscillation
period $\sim 1,000$ s, but after $\sim 5,000$ s the loop settled into
a quasi-steady state with a maximum upward velocity $V < 5$ km/s and
with a coronal gas pressure $P\sim 10^{-1}$ ergs/cm$^3$. This flow
implied a substantial enthalpy flux $5PV/2 = 10^5$ ergs/cm$^3$/s,
which was sufficient to balance the excess radiative losses in the
dipped portion of the loop, so that a steady state became possible.

We let the simulation run for over 30,000 s after the condensation
first appeared in order to verify that a true steady state was
achieved.  There was no significant change in the velocity profile
after the first $\sim 5,000$ s. The evolution consisted of a very slow
growth of the condensation. In principle, the condensation should
expand until it encompasses most of the dipped portion of the loop in
Figure 1, but this would require a time scale much longer than typical
prominence lifetimes. The key point is that the total mass $M$ in the
condensation must increase as $\exp(z/H_g)$, where $z$ is the vertical
depth of the condensation and $H_g$ is the gravitational scale height
$\sim 200$ km for typical prominence temperatures.  Hence, for a fixed
velocity, the condensation's expansion rate slows down exponentially
rapidly with time, so we don't expect condensations to become
significantly larger than their scale heights.

The evolution calculated by our simulation agrees very well with
prominence observations. From the time that the midpoint temperature
first dropped below 30,000 K, the condensation required 500 s to reach
a half-width of 1,000 km and approximately 1,000 s to reach a half
width of 2,000 km, as shown in Figures 5 and 6.  These numbers are
typical of the time scales quoted for the formation time of prominence
condensations (\cite{Engvold76,Zirker89}).  Furthermore, the
condensation growth slowed down abruptly after it reached a half-width
of order 1,000 km; the half-width was only 5,000 km after 30,000 s.

\section{Discussion}

The results described above have several interesting implications for
coronal heating theories and for observations. The most important
feature of our model is that the heating must be localized near the
chromospheric footpoints of a loop; however, this localization is not
severe, since we used a spatial scale of 10,000 km for the heating. On
the other hand, our results do imply that any heating theory in which
the energy input is predicted to occur uniformly, or localized near
the loop top, would be incompatible with the observation of coronal
condensations. Of course, even if the heating were uniform or
localized near the top, transient condensations could still form if
the heating underwent strong temporal variations as in the case of
post-flare H$_\alpha$ loops.

Another implication of the model is that a prominence loop should be
cooler and less dense (except in the condensation itself) than a
standard, uniformly-heated coronal loop with the same total
heating. The reason is that, in the standard coronal loop model, the
total energy input must be balanced by the total radiation emitted
({\it e.g.,} \cite{Rosner78,Vesecky79}), but in our prominence
solution, some of the energy input goes into driving the mass flow
which lifts material against gravity and powers the extra radiation
losses at the transition region of the condensation. The situation is
physically similar to comparing a closed coronal loop in a quiet
region and an open fluxtube in a coronal hole. Even if the heating in
each region has the same magnitude, the quiet-region static loop will
have a substantially higher temperature and density than the
coronal-hole steady-flow fluxtube and, hence, will appear much
brighter in X-rays.

To verify this claim we ran a simulation in which the localized
heating $E_1$ was spread out uniformly over the whole loop length, so
that we would obtain a solution with no condensation.  The resulting
temperature and density profiles looked very similar to the first
curves in Figures 2 and 3, except that the temperature maximum at the
midpoint had a value over $2.6\times 10^6$ K and the density was over
$2.6 \times 10^9$ cm$^{-3}$.  These numbers are to be compared to the
coronal temperatures and densities of the quasi-steady state of the
condensing loop above, $ 1.0 \times 10^6$ K and $1.5 \times 10^9$
cm$^{-3}$ respectively. Even though the total heating in both cases
was identical, the ``coronal loop'' (uniform heating case) had a
coronal density almost two times larger than the ``prominence loop''
(localized heating case) and, therefore, would have an X-ray
brightness four times larger. We conjecture that this effect may play
a role in the origin of the dark cavities that are often seen to
surround prominences.  In our model the region of dipped magnetic flux
tubes threading a prominence can be thought of as forming a local
coronal-hole region, in which plasma is undergoing a quasi-steady flow
into the prominence mass. Even though the low density of the cavity
would make such flows difficult to detect, it should be informative to
search for them with SOHO and TRACE.

Our model predicts other flows which may be detectable by SOHO, as
well. The key point is that if coronal heating has the type of spatial
variation assumed here, then condensations should form in sufficiently
high loops even if they don't have dips. If the loop does have a dip
then a quasi-steady state is possible in which the condensation
settles to the bottom of the dip, but if the loop has no dip, we
believe that no steady state is possible, and the loop must be
constantly dynamic. We propose that this is the origin of the
continuous dynamics observed in very high active region loops with the
CDS instrument on SOHO (\cite{Brekke97}). Further simulations with
non-dipped loop geometries are clearly needed.

The main conclusion of this paper is that our model successfully
reproduces the observed time scales for condensation formation. Not
only does the condensation in the simulation reach a size of a few
thousand km in tens of minutes, but equally important, the
condensation stops growing significantly once it reaches this
size. Both results are necessary in order to agree with observations
of prominence knots.  We believe, therefore, that loss of thermal
equilibrium driven by a spatially localized coronal heating is the
answer to the question: How does mass condense in the corona?

\acknowledgments

This work has been supported in part by NASA and ONR.

\vfill\eject

\begin{figure}
\caption{ Geometry of our model loop. Both horizontal and vertical
distances are measured in units of 10,000 km. \label{fig1}}
\end{figure}

\begin{figure}
\caption{ Temperature as a function of position along the loop
for four times in the evolution. The temperature is measured in units
of $10^6$ K and distance in units of 10,000 km. The solid curve
corresponds to $t = 0$, the time when the localized heating is turned
on. The dashed curve corresponds to $t = 10,000$ s, the dotted curve
to $t = 30,000$ s, and the dash-dot curve to $t = 50,000$ s. \label{fig2}}
\end{figure}

\begin{figure}
\caption{ As in Figure 2, but for number density in units of
$10^9$ cm$^{-3}$. Note that the chromospheric and transition region
portions of the loop are not shown so that
the coronal density structure can be seen clearly. \label{fig3}}
\end{figure}

\begin{figure}
\caption{ As in Figure 2, but for velocity in units of 10 km/s. \label{fig4}}
\end{figure}

\begin{figure}
\caption{ Temperature as a function of position near the loop midpoint
for five times during the growth of the condensation. The temperature
is measured in units of $10^6$ K and distance in units of 10,000
km. The solid curve corresponds to $t = t_0 = 54,000$ s, the time that
the condensation first appeared. The long-dashed curve corresponds to
500 s after $t_0$, the dotted curve to 1,000 s, the dash-dot curve to
10,000 s, and the short-dash curve to 30,000 s after
$t_0$. \label{fig5}}
\end{figure}

\begin{figure}
\caption{ As in Figure 5, but for number density in units of 
$10^{10}$ cm$^{-3}$. \label{fig6}}
\end{figure}

\begin{figure}
\caption{ As in Figure 5, but for velocity in units of 10 km/s
and plotted over the whole loop. \label{fig7}}
\end{figure}

\end{document}